\documentclass[a4paper]{jpconf}
\usepackage{graphicx}
\begin{document}
\title{The two dimensional Hubbard model: \ a theoretical tool for molecular electronics}

\author{Vladan Celebonovic}

\address{Institute of Physics, Pregrevica 118,11080 Zemun-Belgrade,Serbia}

\ead{vladan@ipb.ac.rs}

\begin{abstract}
When speaking about molecular electronics, the obvious question which occurs is how does one study it theoretically. The simplest theoretical model suitable for application in molecular electronics is the two dimensional Hubbard model. The aim of the present paper is to introduce this model, and give some examples of the systems which it can describe. After a short mathematically oriented discussion, it will be shown how to calculate the electrical conductivity of a particular planar system: a rectangular lattice with mutually independent conductivities along the two axes,but without using the $2D$ Hamiltonian. This system could find applications in high Tc studies. It will finally be shown that the electrical conductivity of graphene can be determined not by using the full formalism of the $2D$ Hubbard model, but by a slight reformulation of the Hamiltonian of the $1D$ Hubbard model.\footnote{Lecture given at the 16 Int.School Cond.Matt.Phys.,August 29.- Sept 3,2010.,Varna (Bulgaria). Published: 2010,J.Physics:Conf.Series,{\bf 253},012004}    
\end{abstract}
\section{Introduction}
The electron was discovered in 1897. It was quickly shown that metals contained electrons which were free to move,while insulators did not. In this way the famous problem of metal $\rightarrow$ insulator transitions ($MI$ transitions for short) entered modern physics. Some sucess in analyzing the $MI$ transition was achieved in the early days of quantum mechanics ([1] and references given there). A more general theory of $MI$ transitions was formulated towards the middle of the last century [2]. 

Attempts to develop a microscopic theory of the $MI$ transition were crucial in the development of the so called Hubbard model $HM$ ([3] and later work) . In its mathematically simplest form, the main ingredient of the $HM$ is a collection of atomic orbitals, and the main physical idea is the idea of {\it tight binding}. Expressed in common language, this means that the wave function of an electron is centered on the lattice site of an ion, and that any electron can "hop" one lattice spacing at a time. 

In statistical physics, a model is specified by its Hamiltonian. The Hamiltonian of the HM is given as the sum of two terms, the "free" kinetic term $H_{0}$ and the "interaction" term $H_{I}$
\begin{equation}
	H=H_{0}+H_{I}
\end{equation}  
For an example of a recent discussion on the formulation of the Hubbard model see [4]. In two dimensions, expressed within the second quantisation formalism, the Hamiltonian of the Hubbard model has the following form: 
\begin{eqnarray}
H=-\sum_{ij}\sum_{\sigma}t_{ij}c_{i\sigma}^{+}c_{j\sigma}+\frac{1}{2}\sum_{ijkl}\sum_{\sigma\sigma'}<ij|v|kl>c_{i\sigma}^{+}c_{j\sigma'}^{+}c_{l\sigma'}c_{k\sigma}	
\end{eqnarray}

In this expression,latin indices denote sites on a lattice, $\sigma$ is the spin index, and symbols of the form $c_{i\sigma}^{+}$ denote second quantisation operators creating an electron with spin $\sigma$ at a lattice site $i$. The hopping integral between sites $i$ and $j$ is denoted by $t_{ij}$. Finally, $v$ is the Coulomb potential. 

The Hubbard model may seem relatively simple, judging by the form of its Hamiltonian. However, it is only {\it apparently} simple. Although nearly 60 years have passed since it was proposed, this model has been exactly solved only for the 1D case [5]. In spite of that, the Hubbard model and its applications are still provoking scientific interest. A recent search ( beginning of August 2010 ) with the key words "Hubbard model" gave 1043 papers published in the journal Physica B. A similar search on http://prola.aps.org gave 1613 papers in Phys Rev B and Phys Rev Lett with the same key words in 2000.-2010.   

The relevance of the $2D$ Hubbard model to modern condensed matter physics  follows from the fact that both in "standard" high temperature superconductors, as well as in those containing $Fe$, planar sheets of charges play important roles. 

Calculations are usually rendered easier by introducing the following two approximations in the Hamiltonian:
\[ t_{ij} = \left\{\begin{array}{ll}t & (i,j)\hfill n n  \\                   
                     
                                    0 & otherwise

\end{array}       \right. \]

and
\[ <ij\left|v\right|kl> = \left\{              \begin{array}{ll}                   U & (i=j=k=l)\\                   0 & (otherwise)              \end{array}       \right. \]

where $n n$ denotes nearest neighbors. 

\section{Some mathematical details} 
With these approximations the Hamiltonian of the two dimensional Hubbard model takes the form
\begin{equation}
	H=-t\sum_{<i,j>}\sum_{\sigma}(c_{i,\sigma}^{+}c_{j,\sigma}+h.c)+U\sum_{i}n_{i,\uparrow}n_{i,\downarrow}
\end{equation}
where $h.c$ denotes hermitian conjugation. 

A logical first step in using the $2D$ Hubbard model in theoretical studies of molecular electronics would be finding its general solution. Even with the two approximations mentioned above  the general solution of the $2D$ Hubbard model has not been found.   
The $2D$ Hubbard model can be solved in two particular limiting cases: these are the {\it band limit} and the {\it atomic limit}. 

The {\it band limit} corresponds to the case $U=0$, and the Hamiltonian then has the form 
\begin{equation}
	H_{U=0}=\sum_{\vec{k},\sigma}\epsilon(\vec{k})c_{\vec{k},\sigma}^{+}c_{\vec{k},\sigma}
\end{equation}
and 
    
\begin{equation}
	\epsilon({\vec{k}})=-\sum_{i0} t_{i0}\exp(i \vec{k}\cdot\vec{R_{i}})
\end{equation}

where $\vec{k}$ belongs to the Brillouin zone. In the case of a $2D$ square lattice with lattice step $a$, it can be shown that 
\begin{equation}
	\epsilon(\vec{k}) = - 2 t [\cos(k_{x}a)+\cos(k_{y}a)] 
\end{equation}
\newpage
The {\it atomic limit} of the Hubbard model corresponds to the opposite case: $t=0$ but $U\neq0$. The Hamiltonian then has the form
\begin{equation}
	H_{t=0}=U\sum_{i}n_{i,\uparrow}n_{i,\downarrow}
\end{equation}

As the hopping energy is $t=0$ this limit has no interest for studies of the transport properties of the Hubbard model. The Hubbard model can be solved in each of these two limiting cases separately. Complications arise when attempting to solve it in "intermediary" cases, when both $t\neq0$ and $U\neq0$.

A related problem, with more practical applicability is the calculation of the electrical conductivity of the Hubbard model and,in broader terms, of any physical system whose Hamiltonian is known. These calculations have applications in material science, in studies of materials as different as Q1D organic conductors, high $T_{c}$ superconductors, graphene,...

\subsection{Calculating the conductivity}
Imagine that the Hamiltonian H of a many-body system is at some sufficiently remote moment in the past perturbed by a time-dependent external field $h(t)$ . In real experiments, this external field can be high external pressure, which is at some moment turned on and increases with time. At time $t$ this Hamiltonian can be expressed as $H(t)=H+V(t)$, where $V(t)=h(t)A$, and the symbol $A$ denotes the parameter of the system with which the perturbing external field $h(t)$ is coupled. A common example of $A$ is the particle number density. 

As a consequence of the existence of the perturbation $h(t)$, the system under study is not "isolated" any more. This implies that the average value of the observable represented by the operator $A$ depends on the details of the perturbing field; solving such a problem is a complicated task. This problem can be reduced to the Linear Response Theory if the perturbation $h$ is small enough.

This idea is the basis of the statistical mechanical theory 
of ireversible processes, proposed by Kubo [7]. The aim of this theory is to develop a scheme for the calculation of the kinetic coefficients for quantities such as the electrical and thermal conductivity. Kubo has shown that this calculation can be performed as a calculation of time correlation functions in equilibrium. From the viewpoint of pure theory, Kubo's theory solves the problem - it gives formal expressions for the required physical quantities. However, the expressions it gives are far too complex for application to real materials. 

Another method applicable to the calculation of the kinetic coefficients is the so called "memory function" method, recently reviewed in [8]. This method is a logical continuation of the work by Kubo. It was practically developed in the 1970s, and within this method the electrical conductivity can be calculated as follows:
      \begin{eqnarray}
	\chi_{AB}(z)=<<A;B>>= -i\int_{0}^{\infty}\exp{izt}<[A(t),B(0)]> dt
\end{eqnarray} 

where ${\bf A}={\bf B}=[{\bf j},{\bf H}]$, ${\bf j}$ denotes the current operator, and 

\begin{equation}
	\sigma(z) = i\frac{\omega_{P}^{2}}{4\pi z} [1-\frac{\chi_{z}}{\chi_{0}}]
\end{equation}

The symbol $\omega_{P}$ denotes the plasma frequency, which is given by $\omega_{P}^{2}=4\pi e^{2} n/m$, and where $e,n,m$ are the electron charge, number density and mass respectvely. The current operator is defined as follows [9]:
\begin{equation}
	{\bf j}=\frac{\partial}{\partial t} {\bf P} = i[{\bf H},{\bf P}]
\end{equation}
where 
\begin{equation}
	{\bf P}=\sum_{i}{\bf R_{i}} n_{i} 
\end{equation}
is the polarisation operator. In expression (11) ${\bf R_{i}}$ denotes the position of lattice site $i$, and $n_{i}$ is the number of particles at lattice site $i$. Calculating the electrical conductivity of a $2D$ Hubbard model by the memory function method is, in principle, possible but it would be tedious. Just as an illustration, the current operator would be given by an expression of the form:

\begin{equation}
{\bf j}=i[{\bf H},{\bf P}]=[{\bf H_{0}}+{\bf H_{I}},\sum_{i,j}{\bf R_{i,j}}n_{i,j}]
\end{equation} 
in which the free part and the interaction part of the Hamiltonian would have to be inserted from the Hamiltonian of the 2D HM. Note that in the last expression $R$ and $n$ are labelled by two indexes because of the dimensionality of the lattice.   

The following section of this paper contains a solution to the simplified problem: the calculation of the conductivity of a $2D$ square lattice, with mutually independent conductivities along the axes. 
\section{The conductivity of a square lattice} 
Imagine a $2D$ square lattice of side length $a$. If the lattice sides are denoted by $x$ and $y$, the electric current flowing through such a system can be expressed as

\begin{equation}
	\vec{j}=j_{x}\vec{e_x}+j_{y}\vec{e_{y}}
\end{equation}
where $\vec{e_x}$ and $\vec{e_{y}}$ are unitary vectors of the two lattice axes. The total current is given by 
\begin{equation}
	j=[j_{x}^{2}+j_{y}^{2}]^{1/2}
\end{equation}
As by definition ${\vec j}=\sigma {\vec E}$ where $\sigma$ is the electrical conductivity and ${\vec E}$ the electric field, and assuming that $E_{x}$ = $E_{y}$ = $E$, it finally follows that 
\begin{equation}
	\sigma=[\sigma_{x}^{2}+\sigma_{y}^{2}]^{1/2}
\end{equation}
It is assumed here that the conductivities along the two axes are mutually independent, which means that the conductivity of this system can be exressed using known results for the conductivity of $1D$ correlated electron systems [10]. Full details of the calculation are given in [10]. The electrical conductivity is given by:        
\begin{equation}
\sigma_{R}(\omega_{0})=(1/2\chi_{0})(\omega_{P}^{2}/\pi)[\omega_{0}^{2}-(bt)^{2}]^{-1}(Ut/N^{2})^{2}\times S=p_{F}\times S
\end{equation}
and the symbol $S$ denotes the following function
\begin{eqnarray}
S = 42.49916\times(1+\exp(\beta(-\mu-2t)))^{-2}+78.2557\times
\nonumber\\
(1+\exp(\beta(-\mu+2t\cos(1+\pi))))^{-2}+(bt/(\omega_{0}+bt))\times
\nonumber\\	
(4.53316\times(1+\exp(\beta(-\mu-2t)))^{-2}+
\nonumber\\
24.6448(1+\exp(\beta(-\mu+2t\cos(1+\pi)))))^{-2})
\nonumber\\
\end{eqnarray}
where $\mu$ denotes the chemical potential of the electron gas, determined in [10] as:
\begin{equation}
\mu=\frac{(\beta t)^{6}(ns-1)\left|t\right|}{1.1029+.1694(\beta t)^{2}+.0654(\beta t)^{4}}	
\end{equation}
In equations (15)-(17), $\omega_{0}$ denotes the real part of the frequency, $\beta$ is the inverse temperature, $s$ the lattice constant, $n$ the mean number of electrons per lattice site and $U$ is the on-site Coulomb repulsion. The number of lattice sites is denoted by $N$. The static limit of the susceptibility is denoted by $\chi_{0}$. Equations (15)-(17) give the possibility for investigating the dependence of the conductivity on the temperature $T$, band filling $n$, frequency $\omega_{0}$ and hopping parameter $t$. The symbol $b$ denotes a numerical constant, $b=-1.83879$. 
Applying expressions (15)-(17) is straightforward. For the sake of simplicity, it was assumed that the parameters have the same values along the two axes, meaning that $t_{x}=t_{y}$ etc. The numerical values were chosen as follows: $N=150$, $\chi_{0}=1/3$, $U=4t$, $\omega_{0}=3t$, $\omega_{P}=12t$, $\beta=11600/T$, and the conductivity was normalized to $1$ at $n=0.7$,$t=0.01eV$, $t=100K$.  Examples of results of such calculations are presented in the figures; figures 1 and 2 correspond to $n=0.7$, figure 3 corresponds to $n=0.8$ and figure 4 to $n=1.2$. The figures show that the values of the hopping energy and doping have a drastic influence on the shape of the teperature dependence of the conductivity, as well as on the intensity of the maximum. The existence of such an influence is known in experiments on high $T_{c}$ materials and $Q1D$ organic conductors; see for example [11] and references given there. 

An interesting extension of the results of the present calculation could be obtained by redefining the total current as
\begin{equation}
	\vec{j}=m j_{x}\vec{e_x}+n j_{y}\vec{e_{y}}
\end{equation}
with $m$ and $n$ being any real numbers.
\begin{figure}[h]
\includegraphics[height=12cm]{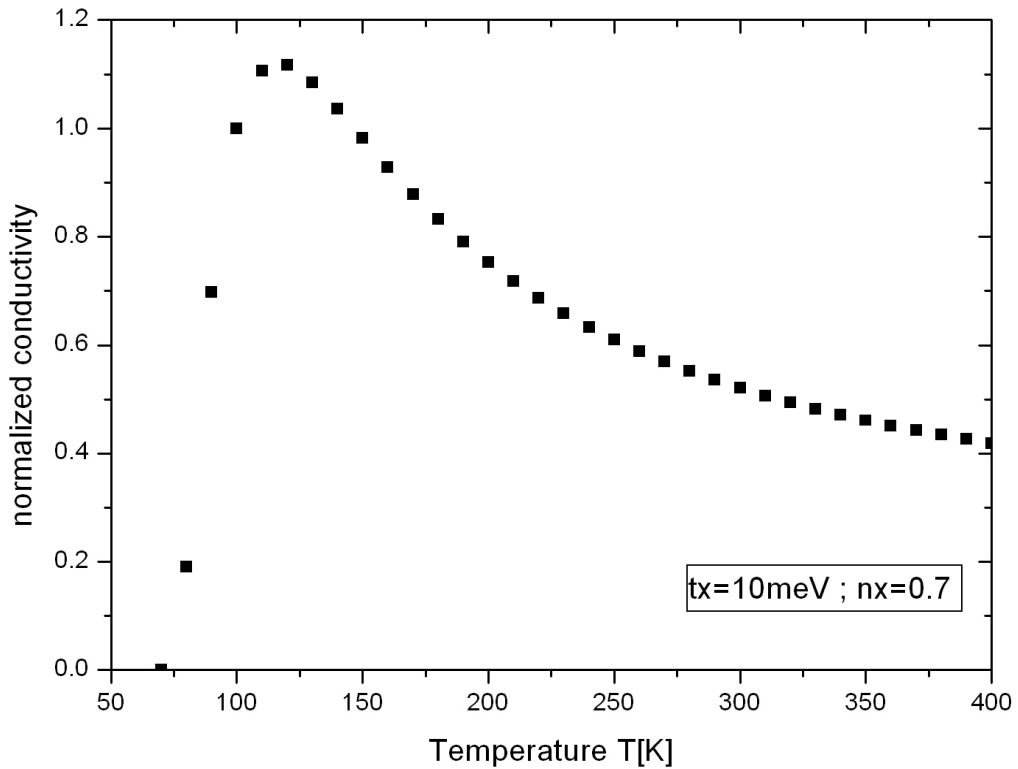}
\caption{Normalized conductivity of a square lattice for $t_{x}=10$ meV and $n=0.7$.}
\end{figure}
\begin{figure}[ht]
\includegraphics[height=12cm]{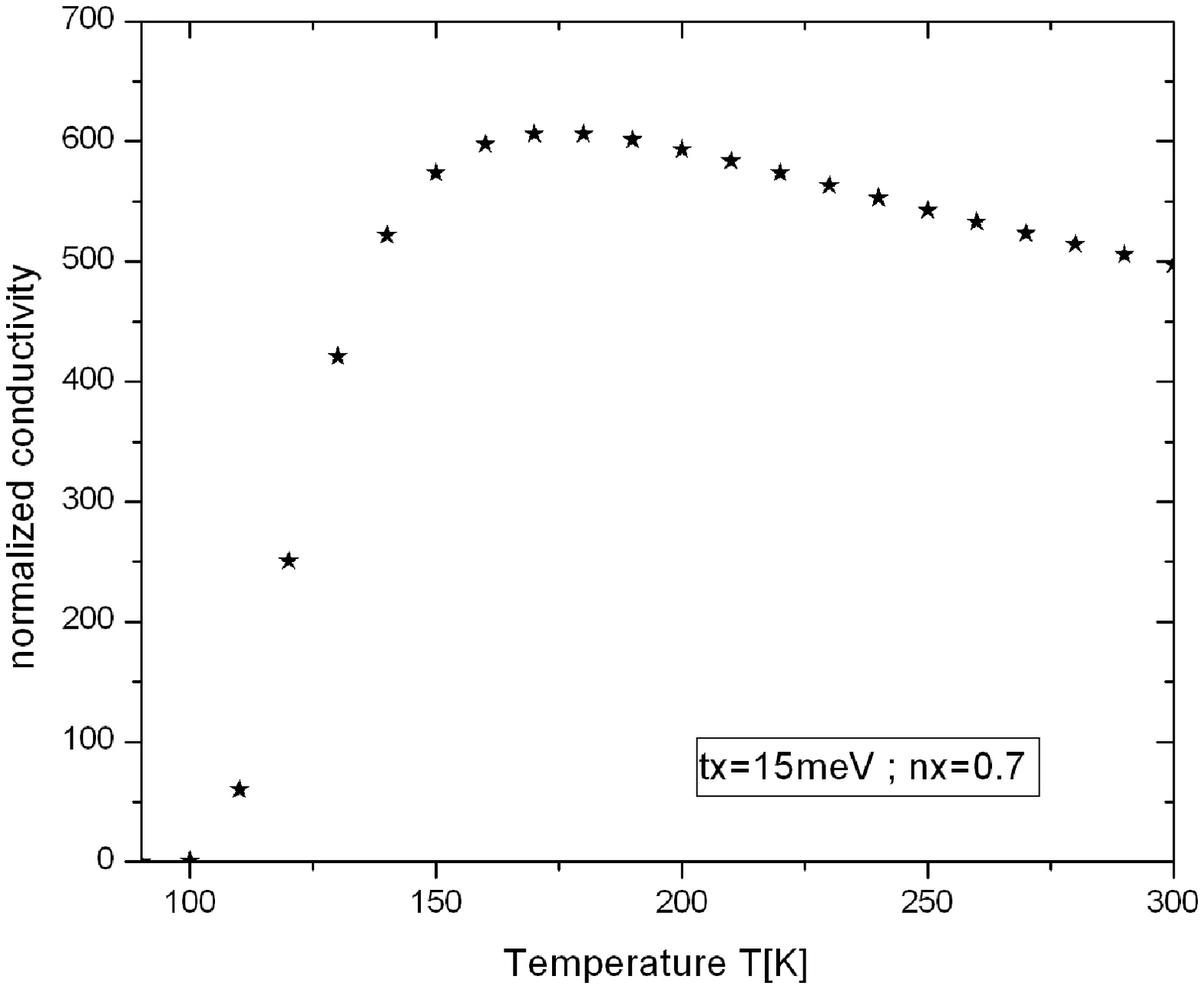}
\caption{Normalized conductivity of a square lattice for $t_{x}=15$ meV and $n=0.7$.}
\end{figure}

\begin{figure}[ht]
\includegraphics[height=12cm]{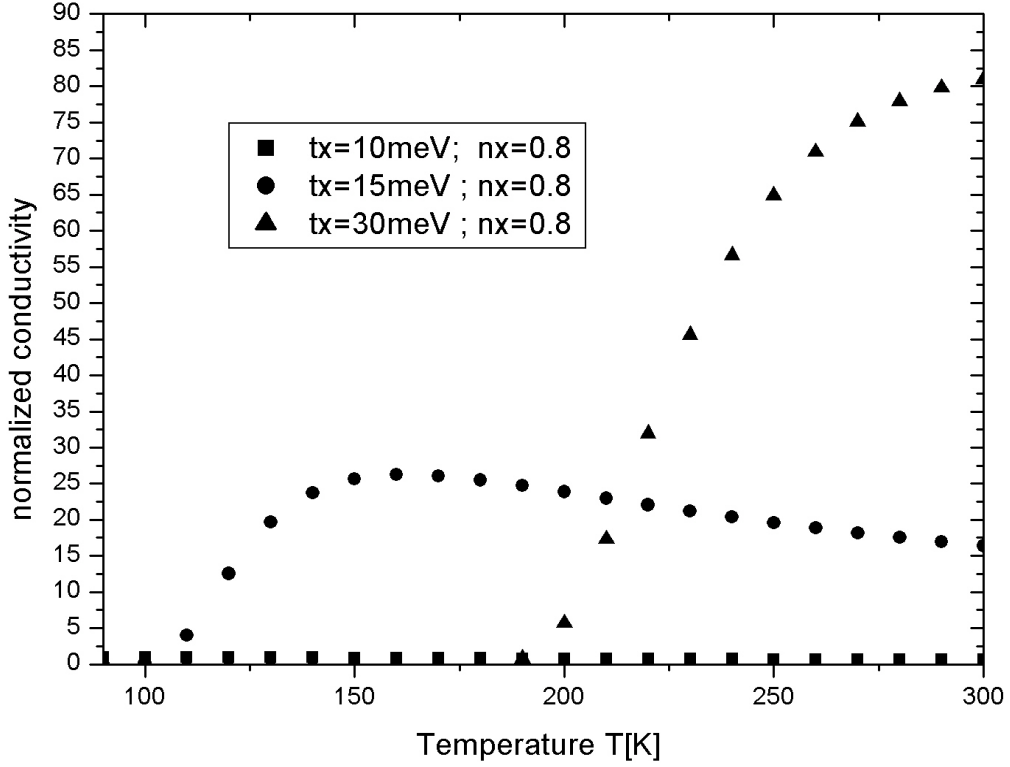}
\caption{Normalized conductivity of a square lattice for 3 values of $t_{x}$  and $n=0.8$.}
\end{figure}

\begin{figure}[ht]
\includegraphics[height=12cm]{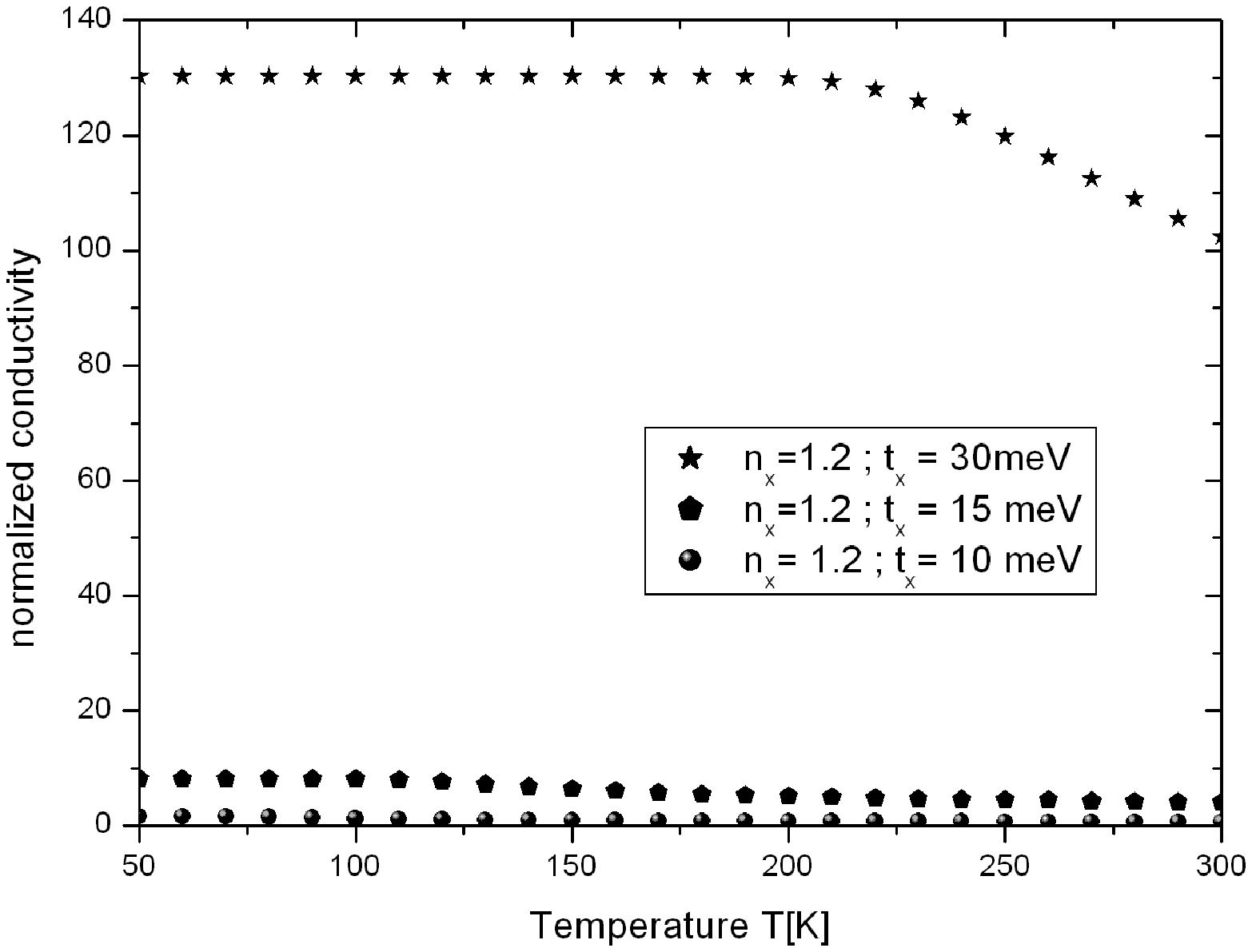}
\caption{Normalized conductivity of a square lattice for $t_{x}$ meV and $n=1.2$.}
\end{figure}

\section{Graphene: is the 2D Hubbard model applicable?} 
It was a "certainty" in statistical physics for more than 70 years that 2D materials cannot exist. It was argued by people of no less scientific stature than Landau or Peierls (quoted in [12]) that such materials would be therodynamically unstable, because a divergent contribution of thermal fluctuations in low dimensional crystal lattices would lead to atomic displacements comparable to atomic spacings. Much to the astonishement to the scientfic community, a $2D$ allotrope of carbon, later named graphene, was discovered in 2004 [13]. Graphene is a monolayer of carbon atoms packed in a honeycomb lattice [14].
\begin{figure}[ht]
\includegraphics[height=13cm]{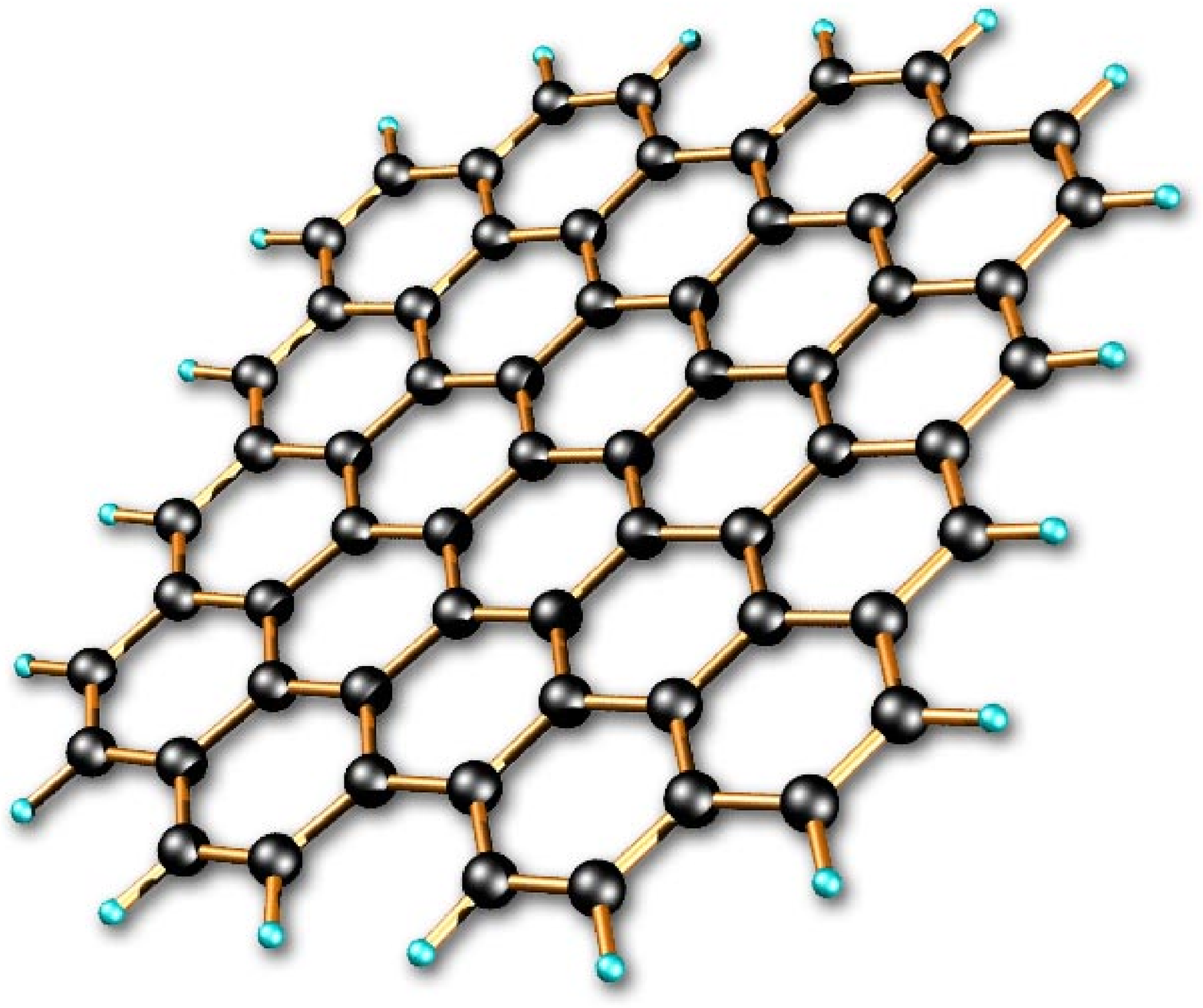}
\caption{The structure of graphene}
\end{figure}
Experiments have shown that graphene is an extremely interesting material. For example, it is the thinnest and strongest known material, and it can support current densities six times greater than copper [15]. It has been shown that electron transport in graphene is governed by the Dirac equation, that its conductivity never becomes zero, and that the charge cariers in graphene are massless Dirac fermions [14]. 
Graphene being planar, the obvious question is whether or not its transport properties can be studied within the Hubbard model, using, for example, the memory function method. Performing such a calculation would be possible by insertion of expression (3) into equations (8)-(10), but it could easily become complex.
There is however a seemingly simpler way to complete such a calculation: use the Hamiltonian of the $1D$ Hubbard model, and simply change the physical meanings of various variables which occur in it. The Hamiltonian of the $1D$ Hubbard model has the following mathematical form:
\begin{equation}
H=-t\sum_{i,\sigma}(c_{i+1\sigma}^{+}c_{i\sigma}+c_{i\sigma}^{+}c_{i+1\sigma})+U\sum_{i}n_{i,\uparrow}n_{i,\downarrow}	
\end{equation}
Symbols of the form $c_{i\sigma}^{+}$ denote second quantisation operators creating an electron with spin $\sigma$ at a lattice site $i$, while $n_{i,\uparrow}$ is the number operator for the number of electrons at an ion on lattice site $i$ having spin up. The calculation of the electrical conductivity using this Hamiltonian and the "memory function" method is discussed in detail in [10]. With application to graphene in view, the Hamiltonian given by equation (19) can be reformulated in the following form
\begin{equation}
	H=t ( \sum_{j,\Delta,\sigma}c_{j,\sigma}^{+}c_{j+\Delta,\sigma}+h.c)+U\sum_{j}c_{j,\sigma}^{+}c_{j,\sigma}c_{j,-\sigma}^{+}
c_{j,-\sigma}
\end{equation}
where  $c_{j\sigma}^{+}$ and $c_{j\sigma}$ are the creation and anihilator operators of electrons in a graphene unit number $j$ and spin $\sigma$, $t$ is the overlap integral between adjacent graphene units, which is determined by overlapping of the wave functions of graphene electrons, $\Delta$ is the vector relating two adjacent units in the hexagonal and $U$ is the energy of Coulomb repulsion of the electrons in the same graphene unit.

From the point of view of mathematical form, equations (19) and (20) are almost identical; this implies that the results of any calculations in which they are used are also going to be of similar form. There exist, however, differences in the physical nature of the quantities which occur in them. For example, in a $1D$ problem, electrons "hop" between lattice sites, while in the case of graphene charge carriers hop between graphene units. 

At the time of this writing (mid August 2010) attempting to apply the $2D$ Hubbard model to the conductivity of graphene is an ongoing calculation. The commutator $[j,H]$ has been calculated and it has the same mathematical form as the corresponding commutator in the $1D$ case. Details are slowly "coming in" and will be discussed elsewhere.
 
\section{Conclusions}  
In this paper the Hubbard model in $2D$ was introduced and discussed to some extent. No attempt was made of a general discussion of the model, but attention was focused on aspects relevant for its transport properties. This aspect is useful for theoretical studies in molecular electronics, which is the topic of this meeting. A method capable of providing a full general solution for the electrical conductivity, called the memory function method, was briefly discussed. The general expression for the conductivity was not discussed. 

However, the conductivity of a particular system: a square lattice with mutually independent conductivities along the axes was discussed to some extent. The temperature dependence, as well as the dependence on doping, were shown on several numerical examples, calculated for a set of arbitrarily chosen lattice parameters. Possibilities for rendering this calculation more complex, and therefore more physically interesting, were also indicated. 

The paper ends with a brief discussion of graphene and a possibility of using the Hubbard model in the calculation of its conductivity. It is pointed out that by slightly reformulating the physical meaning of various parameters in the $1D$ Hubbard model, it should be possible to use it in calculating the conductivity of graphene. Some details of this calculation have been performed, the results are encouraging, but the full calculation is far from being finished. 

An interesting detail has emerged after this paper was completed. Namely, the idea of trying to use the $2D$ Hubbard model was here proposed by purely theoretical reasoning: this model gives good results for the conductivity of $1D$ correlated electron systems, so it seemed a logical extension to try it in the $2D$ case. Some experiments studying the influence of electron-electron interaction on the low temperature conductivity of graphene have just been published, which can be interpreted as an argument in favour of the purely theoretical proposal made in the present paper [16].

\ack
The preparation of this paper was financed through the research project 141007 funded by the Ministry of Science and Technological Development of Serbia.




\end{document}